\documentclass[sigconf]{acmart}

\usepackage{booktabs} 
\usepackage{enumitem}

\setcopyright{rightsretained}

\acmDOI{10.475/123_4}

\acmISBN{123-4567-24-567/08/06}

\acmYear{2017}
\copyrightyear{2017}

\acmPrice{00.00}

\begin{document}
\title{Exploring Content-based Artwork Recommendation with Metadata and Visual Features}

\author{Pablo Messina}
\affiliation{%
  \institution{Pontificia Universidad Catolica de Chile}
  \city{Santiago}
  \state{Chile}
}
\email{pamessina@uc.cl}

\author{Vicente Dominguez}
\affiliation{%
  \institution{Pontificia Universidad Catolica de Chile}
  \city{Santiago}
  \state{Chile}
}
\email{vidominguez@uc.cl}

\author{Denis Parra}
\affiliation{%
  \institution{Pontificia Universidad Catolica de Chile}
  \city{Santiago}
  \state{Chile}
}
\email{dparra@ing.puc.cl}

\author{Christoph Trattner}
\affiliation{%
  \institution{Modul University}
  \city{Vienna}
  \state{Austria}
}
\email{christoph.trattner@modul.ac.at}

\author{Alvaro Soto}
\affiliation{%
  \institution{Pontificia Universidad Catolica de Chile}
  \city{Santiago}
  \state{Chile}
}
\email{asoto@ing.puc.cl}

\renewcommand{\shortauthors}{Messina et al.}

\begin{abstract}
Compared to other areas, artwork recommendation has received little attention, despite the continuous growth of the artwork market. Previous research has relied on ratings and metadata to make artwork recommendations, as well as visual features extracted with deep neural networks (DNN).
However, these features have no direct interpretation to explicit visual features (e.g. brightness, texture) which might hinder explainability and user-acceptance.

In this work, we study the impact of artwork metadata as well as visual features (DNN-based and attractiveness-based) for physical artwork recommendation, using images and transaction data from the \textit{UGallery} online artwork store. Our results indicate that: (i) visual features perform better than manually curated data, (ii) DNN-based visual features perform better than attractiveness-based ones, and (iii) a hybrid approach improves the performance further.
Our research can inform the development of new artwork recommenders relying on diverse content data.

\end{abstract}

%
%


\maketitle


\vspace{-5mm}
\section{Introduction}
Compared to markets affected by 2008's financial crisis, online artwork sales are booming due to social media and new consumption behavior of millennials. Online art sales reached \$3.27 billions in 2015, and at the current grow rate, they will reach \$9.58 billion by 2020 \cite{forbes1}. 
Notably, although many online businesses utilize recommendation systems to boost their revenue, online artwork recommendation has received little attention compared to other areas such as movies \cite{amatriain2013mining} or music \cite{celma2010music}. 
Previous research has shown the potential of personalized recommendations in the arts domain, such as the CHIP project \cite{aroyo2007personalized}, that implemented a personalized recommendation system for the Rijksmuseum. More recently, He et al. \cite{he2016vista} used pre-trained deep neural networks (DNN) for recommendation of digital art, obtaining good results. Unfortunaly, their method is not applicable for the physical artwork problem as the method assumes that the same item can be bought over and over again. Hence their work only works under the collaborative filtering assumption and also did not investigate explicit visual features nor metadata.

\textbf{Objective}. In this paper, we investigate the impact of different features for recommending physical artworks. In particular, we reveal the utility of artwork metadata, latent (DNN) and explicit visual features extracted from images. We address the problem of artwork recommendation with positive-only feedback (user transactions) over \emph{one-of-a-kind} items, i.e., only one instance of each artwork (paintings) is available in the dataset.


\textbf{Research Questions}. Our work was driven by the following research questions: \emph{RQ1}. How do manually-curated metadata perform compared to visual features?, \emph{RQ2}. How do latent visual features from pre-trained DNNs and explicit visual features perform and compare to each other?, and \emph{RQ3}. Do feature combinations provide the best recommendation performance?

\textbf{Contributions}. Our work makes a contribution to the unexplored problem of recommending physical artworks. We run simulated experiments with real-world transaction data provided by a popular online artwork store based in USA named \textit{UGallery}\footnote{http://www.UGallery.com/}. We also introduce a hybrid artwork recommender which exploits all features at the same time. Our results indicate that visual features perform better than manually-curated metadata. In addition, we show that DNN features work better than explicit attractiveness-based visual features. 


\begin{figure*}[t!]
    \centering
\scalebox{0.85}{
     \includegraphics[width=\textwidth]{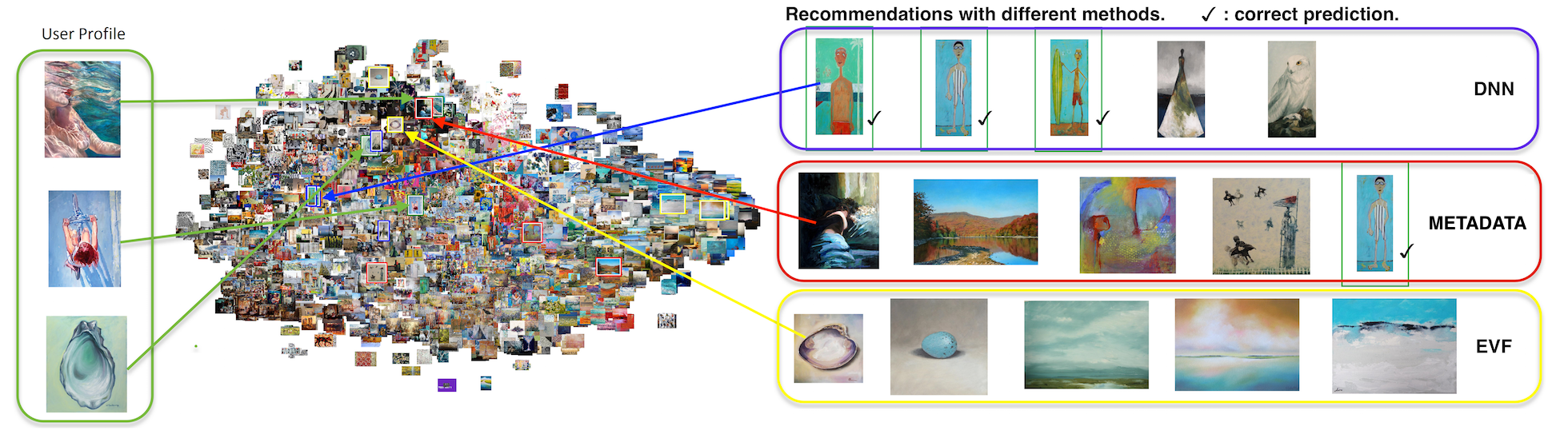}   
    }
    \vspace{-4mm}    
    \caption{t-SNE map of the DNN image embedding displaying paintings of an anonymized user profile (left, green), and recommendations with three methods: DNN (blue), Metadata (red) and EVF (yellow). Check marks indicate correct predictions.}
    \label{fig:ugallery-embedding}
\vspace{-1em}    
\end{figure*}

\vspace{-3mm}
\section{Problem Description}
\label{sec:recproblem}

The online web store \textit{UGallery} supports young and emergent artists by helping them to sell their artworks over their online plattform. To help users of the plattform to explore the vast amount of artworks more efficiently, they are currently investigating with us the possibility of top-n content-based recommendation methods within the plattform 
exploiting features such as artwork metadata, implicit and explicit visual features.

\section{Dataset} UGallery provided us with an anonymized dataset of $1,371$ users, $3,490$ items and $2,846$ purchases (transactions) of paintings, where all users have made at least one transaction. In average, each user has bought 2-3 items 
in the latest years\footnote{Our collaborators at UGallery requested us not to disclose the exact dates when the data was collected.}.


\textbf{Metadata}. Artworks in the \textit{UGallery} dataset were manually curated by experts. In total, there are five attributes: color (e.g. red, blue), subject (e.g. sports, travel), style (e.g. abstract, surrealism), medium (e.g. oil, acrylic), and mood (e.g. energetic, warm).

\textbf{Visual Features}. For each image representing a painting in the dataset we obtain features from an AlexNet DNN \cite{krizhevsky2012imagenet}, which outputs a vector of 4,096 dimensions. We also obtain a vector of explicit visual features of attractiveness, based on the work of San Pedro et al. \cite{sanpedro2009}: brightness, saturation, sharpness, entropy, RGB-contrast, colorfulness and naturalness.

\vspace{-3mm}
\section{Experimental Setup \& Results}
\textbf{Recommendation Methods}. We compare five methods based on the features used: (1) \emph{Metadata}: features based on the metadata of the items previously bought by the user, (2) \emph{DNN}: features from images using the AlexNet DNN \cite{krizhevsky2012imagenet}, (3) \emph{EVF}: Explicit visual features based on attractiveness of the images \cite{sanpedro2009}, (4) \emph{Hyb (DNN + EVF)}: hybrid model using DNN and EVF features, and (5) \emph{Hyb (DNN + EVF + Metadata)}: hybrid model using DNN, EVF and metadata. For the hybrid recommendations, we combine scores of different sources using the BPR framework \cite{rendle2009bpr}. In Figure \ref{fig:ugallery-embedding} we see, for instance, a user profile at the left side, besides the image embedding based on features from AlexNet DNN, and then recommendation obtained by three different methods.

\textbf{Evaluation}. Our protocol is based on the one as introduced by Macedo et al. \cite{macedo2015context} to evaluate recommender system accuratly in a temporal manner. We attempt to predict the items purchased in every transaction, where the training set contains all the artworks previously bought by a user just before making the transaction to be predicted. Users who have purchased exactly one artwork were  remove as their would be no training instance available. 
\textit{Metrics}. As suggested by Cremonesi et al. \cite{Cremonesi2010} for top-n recommendations, we used $recall@k$ and $precision@k$, as well as nDCG \cite{manning2008introduction}.

\textbf{Results}. Table \ref{tab:results} presents the results, which can be summarized as follows: (1) Visual features outperform metadata features. This result is a quite positive finding as manually crafted metadata costs time and money,  (2) visual features obtained from the AlexNet DNN perform better than those based on explicit visual features. Although this result shows that DNNs do again a remarkable job in this domain, we are not too happy about it. Features obtained from an DNN such as AlexNet are latent, i.e., we cannot interpret them directly and we can not use them to explain the recommendations made  \cite{verbert2013visualizing}. Finally, (3) our experiments reveal that the hybrid method performs even best. 

\begin{table}[t!]
\centering
\caption{Results of the simulated recommendation experiment.}
\label{tab:results}
\vspace{-5px}
\scalebox{0.7}
{
  \begin{tabular}{@{\extracolsep{0pt}}lrrrrrr}
    \hline
     name                      &   ndcg@5 &   ndcg@10 &   rec@5 &   rec@10 &   prec@5 &   prec@10 \\
    \hline
Hyb(DNN+EVF+Metadata)  &   \textbf{.0841} &    \textbf{.0990} &  \textbf{.1119} &   \textbf{.1560} &   \textbf{.0279} &    \textbf{.0195} \\
     Hyb(DNN+EVF)       &   .0753 &    .0934 &  .0965 &   .1492 &   .0235 &    .0186 \\
     DNN                       &   .0810 &    .0968 &  .1052 &   .1525 &   .0269 &    .0195 \\
     EVF                       &   .0370 &    .0453 &  .0585 &   .0826 &   .0152 &    .0109 \\
     Metadata                &   .0312 &    .0412 &  .0474 &   .0773 &   .0113 &    .0092 \\
    \hline
  \end{tabular}

}
\vspace{-2mm}
\end{table}

\vspace{-3mm}
\section{Conclusions}
In this work we introduce content-based recommendation for physical artworks, comparing manually-curated metadata, AlexNet DNN features, and attractiveness-based visual features. Furthermore, we show that the DNN features outperform the explicit visual features and metadata. In practice this has two implications: First, there is no need to exploit metadata as visual features work better. Second, it will be difficult to provide explanations to users as explicit features work significanly worse than latend features obtain via DNNs. It would be interesting though to investigate, whether this gap can be closed in a real-world experiment. The current investigations are just based on simulations and neglect the user factor, though give a hint towards the performance of the models when no explanations are given.


\bibliographystyle{ACM-Reference-Format}
\bibliography{sigproc}

\end{document}